\documentclass[10pt,prc,aps]{revtex4}   
\usepackage{graphicx}
\tighten
\begin{document}
\draft
\title{                                                     
Contribution of isovector mesons to the symmetry energy in a     
 microscopic model                
 }              
\author{            
 Francesca Sammarruca}      
\affiliation{                 
 Physics Department, University of Idaho, Moscow, ID 83844-0903, U.S.A  } 
\date{\today} 
\email{fsammarr@uidaho.edu}
\begin{abstract}
We examine the potential energy contributions to the symmetry energy 
(in the parabolic approximation) arising from the isovector 
mesons, $\pi$, $\rho$, and $\delta$. 
The significance of a microscopic model which incorporates all     
important mesons is revealed.                                                      
In particular, we demonstrate 
the importance of the pion for a realistic investigation of 
isospin-sensitive systems. 

\end{abstract}
\maketitle

\section{Introduction} 
                                                                     
The physics of unstable nuclei is closely related to the equation of state 
(EoS) for isospin asymmetric nuclear matter (IANM).
In fact, applications of IANM are broad, 
ranging from the structure of 
rare isotopes to the properties of neutron stars.                                             
An important quantity that emerges from IANM studies is the so-called symmetry energy.
However, in spite of many recent and        
intense efforts, 
the density dependence of the symmetry energy is not sufficiently constrained by the available data and theoretical predictions 
show considerable model dependence.

Older theoretical studies of IANM can be found in Refs.~\cite{BC68,Siem}. 
Interactions adjusted to fit properties of finite nuclei, such as those based
on the non-relativistic Skyrme Hartree-Fock theory \cite{B+75} or the relativistic mean field
theory (see, for instance, Ref.~\cite{ST94}), have been used to extract 
 phenomenological EoS. A review of Skyrme interactions, particularly popular for nuclear structure applications, can be found in 
Ref.~\cite{Stone}. 
Variational calculations of asymmetric matter were reported in Refs.~\cite{12,APR}, 
whereas 
extensive microscopic work with IANM was undertaken by Lombardo and collaborators \cite{Catania1,Catania2} within the Brueckner-Hartree-Fock (BHF) approach. 
Dirac-Brueckner-Hartree-Fock (DBHF) 
calculations of IANM properties were performed by the Oslo group \cite{Oslo},          
the Idaho group \cite{AS03,FS10}, and by Fuchs and collaborators \cite{Fuchs}. 

In this paper, we concentrate on the role of the isovector mesons for the symmetry
energy. The latter is defined from an expansion of the energy/nucleon in terms of 
the isospin asymmetry parameter. In the parabolic approximation, 
it is simply the difference between the energies per particle in neutron matter
and symmetric nuclear matter, see next section. 
Physically, 
it represents the energy ``price" a nucleus must pay for being 
isospin asymmetric.                           

The isovector mesons and their impact on the symmetry energy have been discussed in the literature,
particularly, in the context of mean field approaches, both relativistic and 
non-relativistic (see, for instance, Ref.~\cite{DiToro} for an extensive review 
on reaction dynamics with exotic nuclei based on effective interactions based on 
quantum hadrodynamics (QHD)).  
Lately, considerable interest has developed around the symmetry potential, 
which arises from the difference between neutron and proton single-particle potentials  
in isospin asymmetric matter.
With regard to that issue, it is interesting to recall that, 
in relativistic mean field approaches, introduction of the 
isovector scalar meson (the $\delta$ or $a_0$) is reported to invert the sign of the 
splitting between the masses of the neutron and the proton in neutron-rich
matter \cite{Rizzo}. 

Furthermore, in approaches based on QHD, such as those originally proposed by 
Walecka and collaborators \cite{Wal74,Serot1,Serot2}, the dynamical degrees of
freedom are essentially included through coupling of the nucleons to the isoscalar scalar
$\sigma$ and vector $\omega$ mesons. QHD-I models                                    
of nuclear matter do not     
include the pion, which is perhaps the reason why the contribution of the pion 
to the symmetry energy may have not been discussed in sufficient depth. 
(Note, however, that Walecka's QHD-II model does include both $\pi$ and $\rho$.) 

We will explore the role of all isovector channels for the symmetry energy from the point of view 
of an {\it ab initio} model.                                                    
The main point of the  {\it ab initio} approach is that  mesons are tightly constrained
by the free-space data and their parameters are never readjusted in the medium 
(this is what we mean by ``parameter-free"). 
Furhermore, the contributions from the various mesons are fully iterated, thus 
giving rise to correlation effects. The corresponding predictions can be dramatically
different than those which may 
be produced in first-order calculations. 

This paper is organized as follows: In the next section, we present some facts and
phenomenology about the symmetry energy; then, in Section III, after a brief review 
of our theoretical approach, we focus on exploring the potential energy contributions of the isovector mesons
to the symmetry energy. Our conclusions are summarized in the last section.

\section{Some facts about IANM} 

Asymmetric nuclear matter can be characterized by the neutron density, 
$\rho_n$, and the proton density, $\rho_p$, which are related to their respective
Fermi momenta,    
 $k^{n}_{F}$ or $k^{p}_{F}$, by                                                               
\begin{equation}
  \rho_i =\frac{ (k^{i}_{F})^3}{3 \pi ^2} ,   \label{rhonp}   
\end{equation}
with $i=n$ or $p$. 

It is more convenient to refer to the total density
$\rho = \rho_n + \rho_p$ and the asymmetry (or neutron excess) parameter
$\alpha = \frac{ \rho_n - \rho_p}{\rho}$. 
Clearly, $\alpha$=0 corresponds to symmetric matter and 
$\alpha$=1 to neutron matter.                       
In terms of $\alpha$ and the average Fermi momentum, $k_F$, related to the total density in the usual way, 
\begin{equation}
  \rho =\frac{2 k_F^3}{3 \pi ^2} ,   \label{rho}   
\end{equation}
the neutron and proton Fermi momenta can be expressed as 
\begin{equation}
 k^{n}_{F} = k_F{(1 + \alpha)}^{1/3}            \label{kfn}
\end{equation}
and 
\begin{equation}
 k^{p}_{F} = k_F{(1 - \alpha)}^{1/3} ,            \label{kfp} 
\end{equation}
 respectively.

Expanding 
the energy/particle in IANM  with respect to the asymmetry parameter yields
\begin{equation}
e(\rho, \alpha) = e_0({\rho}) + \frac{1}{2} \Big (\frac{\partial ^2 e(\rho)}{\partial \alpha ^2}\Big )_{\alpha=0}\alpha ^2 +{\cal O}(\alpha ^4) \; , \label{exp}  
\end{equation}
where the first term is the energy/particle in symmetric matter and 
the coefficient of the quadratic term is identified with the symmetry energy,
$e_{sym}$. In the Bethe-Weizs{\" a}cker formula for the nuclear binding energy, it represents the amount of binding a nucleus has 
to lose when the numbers of protons and neutrons are unequal.                                             
To a very good degree of approximation, one can write                   
\begin{equation}
e(\rho, \alpha) \approx e_0({\rho}) + e_{sym}(\rho)\alpha ^2 \;  . \label{e}                    
\end{equation} 
The symmetry energy is also closely related to 
the neutron $\beta$-decay in dense matter, whose threshold depends on the proton fraction. 
A typical value for $e_{sym}$               
at nuclear matter density ($\rho_0$) is 30 MeV, 
with theoretical predictions spreading approximately between 26 and 35 MeV.
The effect of a term of fourth order in the asymmetry parameter (${\cal O}(\alpha ^4)$) on the bulk properties of neutron stars 
is very small, although it may impact the proton fraction at high density. 
More generally, 
non-quadratic terms are usually associated with isovector pairing, which is a surface effect and thus vanishes
in infinite matter \cite{Steiner}. 

Equation~(\ref{e}) displays a convenient separation between the symmetric and aymmetric parts of the EoS, 
which facilitates the identification of observables that, for instance, may be sensitive mainly to the 
symmetry energy. Presently, research groups from GSI \cite{GSI}, MSU \cite{Tsang}, Italy \cite{Greco}, France \cite{IPN}, and 
China \cite{China2,China4} 
are investigating the density dependence of the symmetry energy through heavy-ion collisions. 
Based upon recent results,            
these investigations appear to agree reasonably well on the following parametrization 
of the symmetry energy: 
\begin{equation}
e_{sym}(\rho) = 12.5 \, MeV \Big (\frac{\rho}{\rho_0}\Big )^{2/3} +                       
17.5 \, MeV \Big (\frac{\rho}{\rho_0}\Big )^{\gamma_i},                   \label{es} 
\end{equation} 
where the first term is the kinetic contribution and 
 $\gamma_i$ (the exponent appearing in the potential energy part) is found to be between 0.4 and 1.0. 
Naturally, there are uncertainties associated with all transport models. 
Recent constraints from MSU \cite{Tsang} were extracted from simulations of $^{112}$Sn
and $^{124}$Sn collisions with an Improved Quantum Molecular Dynamics transport model and are 
consistent with isospin diffusion data and the ratio of neutron and proton spectra.

Typically, parametrizations like the one given in Eq.~(\ref{es}) are valid 
at or below the saturation density, $\rho_0$. Efforts to constrain the behavior of the symmetry energy
at higher densities 
are presently being pursued through observables such as $\pi ^-/\pi^+$ ratio, 
$K ^+/K^0$ ratio, neutron/proton differential transverse flow, or nucleon elliptic flow \cite{Ko09}.

\section{The role of isovector mesons} 

\subsection{Review of the theoretical approach}

As stated in the Introduction, 
the starting point of our many-body calculation is a realistic NN interaction which is then applied in the 
nuclear medium without any additional free parameters. 
Thus the first question to be confronted concerns the choice of the ``best" NN interaction. 
After the development of QCD and the understanding of its symmetries,  
chiral effective theories \cite{chi,ME11} were developed as a way to respect the 
symmetries of QCD while keeping the degrees of freedom (nucleons and pions) typical of low-energy nuclear physics. However, 
chiral perturbation theory (ChPT)
has definite limitations as far as the range of allowed momenta is concerned. 
For the purpose of applications in dense matter, where higher and higher momenta become involved     
with increasing Fermi momentum, NN potentials based on ChPT are unsuitable.       

Relativistic meson theory is an appropriate framework to deal with the high momenta encountered in dense
matter. In particular, 
the one-boson-exchange (OBE) model has proven very successful in describing NN data in free space 
and has a good theoretical foundation. 
Among the many available OBE potentials, some being part of the ``high-precision generation" \cite{pot1,pot2}, 
we seek a momentum-space potential developed within a relativistic scattering equation, such as the 
one obtained through the Thompson \cite{Thom} three-dimensional reduction of the Bethe-Salpeter equation \cite{BS}. 
Furthermore, we require a potential that uses 
the pseudovector coupling for the interaction of nucleons with pseudoscalar mesons. 
With these constraints in mind, 
as well as the requirement of a good description of the NN data, 
Bonn B \cite{Mac89} is a reasonable choice.                                                     
The mesons included are the pseudoscalar $\pi$ and $\eta$, the scalar $\sigma$ and 
$\delta$, and the vector $\rho$ and $\omega$.

As our many-body framework, we choose the Dirac-Brueckner-Hartree-Fock  approach.
We will now review the main aspects of our approach and the various approximations 
we perform through the application of the DBHF procedure.

The main strength of the DBHF approach is its inherent ability to account for important three-body forces (TBF) 
through its density dependence. 
These are the TBF originating from virtual excitation of a nucleon-antinucleon pair, 
known as ``Z-diagram".
The characteristic feature of                                     
the DBHF method turns out to be closely related to 
the TBF of the Z-diagram type, as we will argue next. In the DBHF approach, one describes the positive energy solutions
of the Dirac equation in the medium as 
\begin{equation}
u^*(p,\lambda) = \left (\frac{E^*_p+m^*}{2m^*}\right )^{1/2}
\left( \begin{array}{c}
 {\bf 1} \\
\frac{\sigma \cdot \vec {p}}{E^*_p+m^*} 
\end{array} 
\right) \;
\chi_{\lambda},
\label{ustar}
\end{equation}
where the nucleon effective mass, $m^*$, is defined as $m^* = m+U_S$, with $U_S$ an attractive scalar potential.
(This will be derived below.) 
It can be shown that both the description of a single-nucleon via Eq.~(\ref{ustar}) and the evaluation of the 
Z-diagram generate a repulsive effect on the energy/particle in symmetric nuclear matter which depends on the density approximately
as 
\begin{equation}
\Delta E \propto  \left (\frac{\rho}{\rho_0}\right )^{8/3} \, , 
\label{delE} 
\end {equation}
and provides the saturating mechanism missing from conventional Brueckner calculations 
\cite{GB87}. 
(Alternatively, explicit TBF are used along with the BHF method in order to achieve a similar result.)

The approximate equivalence of the effective-mass description of Dirac states and the contribution from the Z-diagram 
has a simple intuitive explanation in the observation 
that Eq.~(\ref{ustar}), like any other solution of the Dirac equation,
can be written as a superposition of positive and negative energy solutions. On the other hand, the ``nucleon" in the 
middle of the Z-diagram is precisely a superposition of positive and negative energy states. 
In summary, the DBHF method effectively takes into account a particular class of 
TBF, which are crucial for nuclear matter saturation.

Having first summarized the main DBHF philosophy, 
we now proceed to describe the DBHF calculation of IANM \cite{AS03}. 
In the end, this will take us back to the crucial point of the DBHF approximation, Eq.~(\ref{ustar}). 

We start from the Thompson \cite{Thom} relativistic three-dimensional reduction 
of the Bethe-Salpeter equation \cite{BS}. The Thompson equation is applied to nuclear matter in
strict analogy to free-space scattering and reads, in the nuclear matter rest frame,                 
\begin{eqnarray}
&& g_{ij}(\vec q',\vec q,\vec P,(\epsilon ^*_{ij})_0) = v_{ij}^*(\vec q',\vec q) \nonumber \\            
&& + \int \frac{d^3K}{(2\pi)^3}v^*_{ij}(\vec q',\vec K)\frac{m^*_i m^*_j}{E^*_i E^*_j}
\frac{Q_{ij}(\vec K,\vec P)}{(\epsilon ^*_{ij})_0 -\epsilon ^*_{ij}(\vec P,\vec K)} 
g_{ij}(\vec K,\vec q,\vec P,(\epsilon^*_{ij})_0) \, ,                                   
\label{gij}
\end{eqnarray}                    
where $g_{ij}$ is the in-medium reaction matrix 
($ij$=$nn$, $pp$, or $np$), and the                                      
asterix signifies that medium effects are applied to those quantities. Thus the NN potential, 
$v_{ij}^*$, is constructed in terms of effective Dirac states (in-medium spinors) as explained above. 
In Eq.~(\ref{gij}),                                  
$\vec q$, $\vec q'$, and $\vec K$ are the initial, final, and intermediate
relative momenta, and $E^*_i = \sqrt{(m^*_i)^2 + K^2}$. 
The momenta of the two interacting particles in the nuclear matter rest frame have been expressed in terms of their
relative momentum and the center-of-mass momentum, $\vec P$, through
\begin{equation} 
\vec P = \vec k_{1} + \vec k_{2}       \label{P}    
\end{equation} 
and 
\begin{equation} 
\vec K = \frac{\vec k_{1} - \vec k_{2}}{2} \, .  \label{K}
\end{equation}                    
The energy of the two-particle system is 
\begin{equation} 
\epsilon ^*_{ij}(\vec P, \vec K) = 
e^*_{i}(\vec P, \vec K)+  
e^*_{j}(\vec P, \vec K)   
\label{eij}
\end{equation} 
 and $(\epsilon ^*_{ij})_0$ is the starting energy.
 The single-particle energy $e_i^*$ includes kinetic energy and potential 
 energy contributions (see Eq.~(\ref{spe}) below).                               
The Pauli operator, $Q_{ij}$, prevents scattering to occupied $nn$, $pp$, or $np$ states.            
 To eliminate the angular
dependence from the kernel of Eq.~(\ref{gij}), it is customary to replace the exact
Pauli operator with its angle-average. 
Detailed expressions for the Pauli operator                     
and the average center-of-mass momentum in the case of two different Fermi seas  
can be found in Ref.\cite{AS03}.                              

With the definitions
\begin{equation} 
G_{ij} = \frac{m^*_i}{E_i^*(\vec{q'})}g_{ij}
 \frac{m^*_j}{E_j^*(\vec{q})}             
\label{Gij}
\end{equation} 
and 
\begin{equation} 
V_{ij}^* = \frac{m^*_i}{E_i^*(\vec{q'})}v_{ij}^*
 \frac{m^*_j}{E_j^*(\vec{q})} \, ,        
\label{Vij}
\end{equation} 
 one can rewrite Eq.~(\ref{gij}) as
\begin{eqnarray}
&& G_{ij}(\vec q',\vec q,\vec P,(\epsilon ^*_{ij})_0) = V_{ij}^*(\vec q',\vec q) \nonumber \\[4pt]
&& + \int \frac{d^3K}{(2\pi)^3}V^*_{ij}(\vec q',\vec K)
\frac{Q_{ij}(\vec K,\vec P)}{(\epsilon ^*_{ij})_0 -\epsilon ^*_{ij}(\vec P,\vec K)} 
G_{ij}(\vec K,\vec q,\vec P,(\epsilon^*_{ij})_0) \, ,                                    
\label{Geq}
\end{eqnarray}                    
which is formally identical to its non-relativistic counterpart.

The goal is to determine self-consistently the nuclear matter single-particle potential   
which, in IANM, will be different for neutrons and protons. 
To facilitate the description of the procedure, we will use a schematic
notation for the neutron/proton potential.                                                   
We write, for neutrons,
\begin{equation}
U_n = U_{np} + U_{nn} \; , 
\label{un}
\end{equation}
and for protons
\begin{equation}
U_p = U_{pn} + U_{pp} \, , 
\label{up}
\end{equation}
where each of the four pieces on the right-hand-side of Eqs.~(\ref{un}-\ref{up}) signifies an integral of the appropriate 
$G$-matrix elements ($nn$, $pp$, or $np$) obtained from Eq.~(\ref{Geq}).                                           
Clearly, the two equations above are coupled through 
the $np$ component and so they must be solved simultaneously. Furthermore, 
the $G$-matrix equation and Eqs.~(\ref{un}-\ref{up})  
are coupled through the single-particle energy (which includes the single-particle
potential, itself defined in terms of the $G$-matrix). So we have a coupled system to be solved self-consistently.

Before proceeding with the self-consistency, 
one needs an {\it ansatz} for the single-particle potential. The latter is suggested by 
the most general structure of the nucleon self-energy operator consistent with 
all symmetry requirements. That is: 
\begin{equation}
{\cal U}_i({\vec p}) =  U_{S,i}(p) + \gamma_0  U_{V,i}^{0}(p) - {\vec \gamma}\cdot {\vec p}  U_{V,i}(p) \, , 
\label{Ui1}
\end{equation}
where $U_{S,i}$ and 
$U_{V,i}$ are an attractive scalar field and a repulsive vector field, respectively, with 
$ U_{V,i}^{0}$ the timelike component of the vector field. These fields are in general density and momentum dependent. 
We take             
\begin{equation}
{\cal U}_i({\vec p}) \approx U_{S,i}(p) + \gamma_0 U_{V,i}^{0}(p) \, ,                                            
\label{Ui2}
\end{equation}
which amounts to assuming that the spacelike component of the vector field is much smaller than 
 both $U_{S,i}$ and $U_{V,i}^0$. Furthermore, neglecting the momentum dependence of the scalar and
vector fields and inserting Eq.~(\ref{Ui2}) in the Dirac equation for neutrons/protons propagating in 
nuclear matter,
\begin{equation}
(\gamma _{\mu}p^{\mu} - m_i - {\cal U}_i({\vec p})) u_i({\vec p},\lambda) = 0  \, ,                                                       
\label{Dirac1} 
\end{equation}
naturally leads to rewriting the Dirac equation in the form 
\begin{equation}
(\gamma _{\mu}(p^{\mu})^* - m_i^*) u_i({\vec p},\lambda) = 0  \, ,                                                       
\label{Dirac2} 
\end{equation}
with positive energy solutions as in Eq.~(\ref{ustar}), $m_i^* = m + U_{S,i}$, and 
\begin{equation}
(p^0)^* = p^0 - U_{V,i}^0 (p) \, .                                                                 
\label{p0}
\end{equation}
The subscript ``$i$'' signifies that these parameters are different for protons and
neutrons. 

As in the symmetric matter case \cite{BM84}, evaluating  the expectation value of Eq.~(\ref{Ui2})       
leads to a parametrization of 
the single particle potential for protons and neutrons (Eqs.(\ref{un}-\ref{up})) in terms of the 
constants $U_{S,i}$ and $U_{V,i}^0$ which is given by      
\begin{equation}
U_i(p) = \frac{m^*_i}{E^*_i}<{\vec p}|{\cal U}_i({\vec p})|{\vec p}> = 
\frac{m^*_i}{E^*_i}U_{S,i} + U_{V,i}^0 \; .      
\label{Ui3}
\end{equation}
Also, 
\begin{equation}
U_i(p) =                                                              
\sum_{j=n,p} 
\sum_{p' \le k_F^j} G_{ij}({\vec p},{\vec p}') \; , 
\label{Ui4}
\end{equation}
which, along with Eq.~(\ref{Ui3}), allows the self-consistent determination of the single-particle
potential as explained below. 

The kinetic contribution to the single-particle energy is
\begin{equation}
T_i(p) = \frac{m^*_i}{E^*_i}<{\vec p}|{\vec \gamma} \cdot {\vec p} + m|{\vec p}> =     
\frac{m_i m^*_i + {\vec p}^2}{E^*_i} \; , 
\label{KE}    
\end{equation}
and the single-particle energy is 
\begin{equation}
e^*_i(p) = T_i(p) + U_i(p) = E^*_i + U^0_{V,i} \; . 
\label{spe}
\end{equation}
The constants $m_i^*$ and 
\begin{equation}
U_{0,i} = U_{S,i} + U_{V,i}^0      
\label{U0i} 
\end{equation}
are convenient to work with as they 
facilitate          
the connection with the usual non-relativistic framework \cite{HT70}.                       

Starting from some initial values of $m^*_i$ and $U_{0,i}$, the $G$-matrix equation is 
 solved and a first approximation for $U_{i}(p)$ is obtained by integrating the $G$-matrix 
over the appropriate Fermi sea, see Eq.~(\ref{Ui4}). This solution is 
again parametrized in terms of a new set of constants, determined by fitting the parametrized $U_i$, 
Eq.~(\ref{Ui3}), 
to its values calculated at two momenta, a procedure known as the ``reference spectrum approximation". 
The iterative procedure is repeated until satisfactory convergence is reached.     

Finally, the energy per neutron or proton in nuclear matter is calculated from 
the average values of the kinetic and potential energies as 
\begin{equation}
\bar{e}_{i} = \frac{1}{A}<T_{i}> + \frac{1}{2A}<U_{i}> -m \; . 
\label{ei}
\end{equation}
 The EoS, or energy per nucleon as a function of density, is then written as
\begin{equation}
    \bar{e}(\rho_n,\rho_p) = \frac{\rho_n \bar{e}_n + \rho_p \bar{e}_p}{\rho} \, , 
\label{enp} 
\end{equation}
or 
\begin{equation}
    \bar{e}(k_F,\alpha) = \frac{(1 + \alpha) \bar{e}_n + (1-\alpha) \bar{e}_p}{2} \, . 
\label{eav} 
\end{equation}
Clearly, symmetric nuclear matter is obtained as a by-product of the calculation described above 
by setting $\alpha$=0, whereas $\alpha$=1 corresponds to pure neutron matter.

\subsection{Results} 

\begin{table}                
\centering \caption                                                    
{Potential energy contributions (in MeV) for selected partial waves to the energy of NM and SNM, and their difference. 
The density is equal to 0.185 fm$^{-3}$.
} 
\vspace{5mm}

\begin{tabular}{|c|c|c|c|}
\hline
Partial waves & 
$U_{NM} $ &                                                               
$ U_{SNM}$ &                                                            
$U_{NM} - U_{SNM}$                                 
\\                                                  
\hline
$^1S_0$&-18.71   &-18.75  & 0.042 \\                                                 
$^3P_0$&-1.88    &-1.75   &-0.126 \\                                                 
$^1P_1$& 0       &4.045   &-4.045 \\                                                 
$^3P_1$&20.51    &14.40   & 6.111 \\                                                 
$^3S_1$&0        &-20.29  & 20.29 \\                                                 
$^3D_1$&0        &1.564   & -1.564\\                                                 
$^1D_2$&-4.250   &-2.477  & -1.773\\                                                 
$^3D_2$&0        &-4.360  & 4.360\\                                                 
$^3F_2$&-1.022   &-0.560  &-0.462\\                                                 
$^3P_2$&-12.47   &-7.697  &-4.773\\                                                 
\hline

\end{tabular}
\end{table}

\begin{table}                
\centering \caption                                                    
{ Contributions (in MeV) to the potential energy of SNM from various mesons 
for three different potential models. The density is equal to 0.185 fm$^{-3}$.
} 
\vspace{5mm}

\begin{tabular}{|c|c|c|c|c|c|c|c|c|}
\hline
Potential & 
$\sigma+\omega $ & $\sigma+\omega+\pi$ & $\pi$ & $\sigma+\omega+\pi+\rho$ & $\rho $ &          
$\sigma+\omega+\rho+\delta$ & $\delta$ & All mesons 
\\                                                  
\hline
Bonn B&  -29.82 & -45.89 & -17.08 & -38.69& 7.21 &-35.13 & 3.56 & -34.24 \\ 
Bonn A&  -33.27 & -44.65 & -11.38 & -38.47& 6.18 &-36.90 & 1.57 & -36.15 \\ 
Bonn C&  -23.45 & -45.21 & -21.75 & -38.74& 6.47 &-33.61 & 5.13 & -32.98    
\\ 
\hline

\end{tabular}
\end{table}

\begin{table}                
\centering \caption                                                    
{ As in the previous Table but for NM.                                       
} 
\vspace{5mm}

\begin{tabular}{|c|c|c|c|c|c|c|c|c|}
\hline
Potential & 
$\sigma+\omega $ & $\sigma+\omega+\pi$ & $\pi$ & $\sigma+\omega+\pi+\rho$ & $\rho $ &          
$\sigma+\omega+\rho+\delta$ & $\delta$ & All mesons 
\\                                                  
\hline
Bonn B&  -17.00 & -13.30 & 3.7008 & -12.00& 1.30 &-15.21 &-3.22 & -16.09 \\ 
Bonn A&  -20.10 & -15.50 & 4.60   & -14.00& 1.50 &-15.23 &-1.23 & -16.40 \\ 
Bonn C&  -14.04 & -11.37 & 2.67   & -10.39& 0.98 &-15.48 &-5.11 & -16.05 \\ 
\hline

\end{tabular}
\end{table}

\begin{table}                
\centering \caption                                                    
{The difference between the potential energy contributions (in MeV) to NM and 
SNM from isovector mesons. 
} 
\vspace{5mm}

\begin{tabular}{|c|c|c|c|}
\hline
Potential & 
$U^{\pi}_{NM} - U^{\pi}_{SNM}$ &                                                               
$U^{\rho}_{NM} - U^{\rho}_{SNM}$ &                                                            
$U^{\delta}_{NM} - U^{\delta}_{SNM}$                                 
\\                                                  
\hline
Bonn B&  20.78  & -5.90  & -6.78 \\                                                 
Bonn A&  15.98  & -4.68  & -2.80 \\                                                 
Bonn C&  24.42  & -5.48  & -10.24 \\                                                
\hline

\end{tabular}
\end{table}

In Table I we show the contributions of some major partial waves to the potential
energy of neutron matter (NM) and of symmetric nuclear matter (SNM). The last column
displays their difference, to signify the potential energy contribution 
to the symmetry energy. The chosen density is 0.185 fm$^{-3}$, corresponding to a Fermi
momentum of 1.4 fm$^{-1}$ in SNM and 
 1.76 fm$^{-1}$ in NM (from Eq.~(3) with $\alpha$=1). (Summing up all contributions
and including the kinetic term yields 33.7 MeV, very close to the actual value of 
our symmetry energy at this density.)

We observe that 
spin-triplet waves, particularly $^3S_1$, give the largest contribution. It will be
interesting to revisit this point in conjunction with the role of the $\delta$ meson.
We note that, 
although the contribution of the $\delta$ meson to 
a quantitative nucleon-nucleon (NN) interaction is 
known to be 
relatively small, this meson is a crucial mechanism to fine-tune the S-waves, that is,
$^1S_0$ {\it vs.} 
$^3S_1$. Hence, its importance for isospin dependent phenomena.

In Table II, we show the contributions to the potential energy of SNM from the different
mesons.                                                                                
We show these contributions for potentials A and C as well \cite{Mac89}. The three 
potentials differ mostly in the parameters used for the $\pi NN$ form factor, which has a large impact on the strength of the tensor force, with Bonn A displaying
the weakest tensor force and Bonn C the strongest (as demonstrated by the predicted
D-state probabilities, which are 4.47, 5.10, and 5.53\%  for A, B, and C, respectively.)
These three potentials span the uncertainty in our knowledge of the short-range tensor
force.
Considering all three models will then provide information on how the effects being
examined (namely, the role of the isovector mesons on the symmetry energy) change with
changing tensor force, while maintaining consistency with the free-space NN data. We 
believe that the latter constraint in crucial for a reliable investigation of many-body 
effects. 

\begin{figure}[!t] 
\centering 
\vspace*{-1.0cm}
\hspace*{-1.0cm}
\scalebox{0.50}{\includegraphics{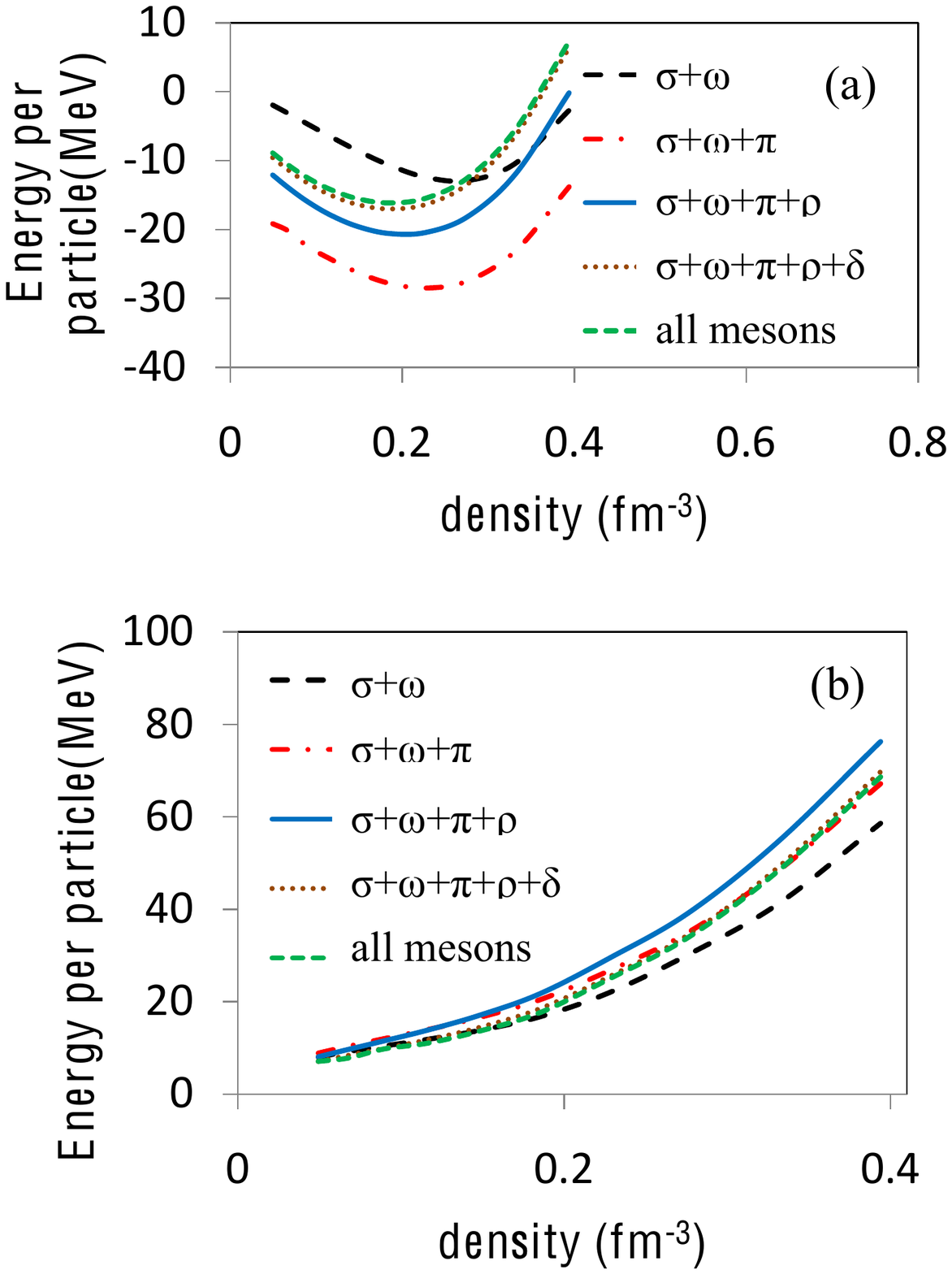}} 
\vspace*{-4.0cm}
\caption{(color online)                                        
Contribution from the various mesons to the equation of state of symmetric matter 
({\bf a}) and neutron matter ({\bf b}). 
} 
\label{one}
\end{figure}

 We start by taking $\sigma$ and $\omega$ together, since a model with only one
of these mesons is entirely meaningless and would produce a completely unrealistic 
correlated wavefunction, especially so with the $\sigma$ alone, due to the absence of any short-range repulsion. The fourth column is the difference between the values shown in 
the third and second columns and represents the contribution from the pion alone.
Notice that this contribution is attractive, as to be expected, recalling that the pion's
tensor potential, $V_t$, generates a large and attractive second-order term,                
$V_t\frac{Q}{E_0-E}V_t$, when iterated in 
the Bethe-Goldstone equation. Consistent with that, this contribution 
is largest with Bonn C, due to its stronger tensor force. 

The contribution of the $\rho$ meson is shown in the sixth column as the difference between
the values in columns five and three. It is considerably smaller than the pion's
and repulsive, since the tensor force generated by the $\rho$ typically reduces the pion's 
tensor force at short range. 
With regard to the $\rho$,                                            
it is useful to recall that the interaction Lagrangian which 
couples vector mesons with nucleons contains both a vector coupling and a tensor
coupling,                                                                           
\begin{equation}
{\cal L}_{NN\rho} = -g_{\rho}\bar{\psi}\gamma_{\mu}{\vec \tau}\psi \cdot \phi^{\mu}_{\rho}
-\frac{f_{\rho}}{4M}\bar{\psi}\sigma_{\mu \nu} {\vec \tau} \psi \cdot 
(\partial ^{\mu}\phi^{\nu}_{\rho}-
\partial ^{\nu}\phi^{\mu}_{\rho}) \; . 
\end{equation}
 These are related to the electromagnetic properties of the nucleon 
in the vector dominance model, where the nucleon couples to the photon {\it via} 
a vector meson. In the framework of the vector dominance model \cite{VDM}, a value 
of 3.7 is obtained for the ratio of the tensor to vector coupling constant,          
$\kappa_{\rho}=f_{\rho}/g_{\rho}$, whereas 
a stronger value of $\kappa_{\rho}$=6.6 was determined from partial-wave analyses \cite{HP}.
In other words, a 
larger value of the $\rho$ tensor coupling as compared to its vector coupling 
is well supported by evidence, 
a fact that is reflected in meson exchange models where, typically, 
the ratio $\kappa_{\rho}$ is about 6.    
Therefore, a Lagrangian density with only a vector coupling for the $\rho$                     
\cite{BP89}, {\it i.~e.} $f_{\rho}$=0, may miss the most important part of how this meson couples to  
the nucleon.     

Back to Table II, 
 the $\delta$ meson is included next, providing a small and positive
contribution. The last column displays the full result, when the pseudoscalar meson
$\eta$ is included as well. 
Table II is more insightful when examined together with Table III. The latter shows the 
same physical quantities as in Table II but for pure neutron matter. 
Here, the contribution of the pion is much smaller and opposite in sign. This is 
due to the absence of the $^3S_1$ partial wave in NM and, 
consequently, the absence of a large part of the attractive second-order tensor
term mentioned above. 
The  effect of the $\delta$ meson in NM is of about the same size as the one observed in SNM
but opposite in sign. 
 This can be easily understood recalling
that the effect of the isovector scalar meson is attractive in $^1S_0$ and repulsive
in $^3S_1$, and that the latter is absent from NM. 
With respect to potential model dependence, 
the size of the effect is largest in model C and weakest in model A. 
Model dependence should be expected, as 
the parameters of the $\delta$ meson are quite different for the three potentials. 

Before proceeding to discuss the symmetry energy, 
we show, for Bonn B, how the various mesons contribute to the energy of symmetric 
nuclear matter, Fig.~1(a), and neutron matter,
Fig.~1(b).                                                                          
From Fig.~1(a), one can see that the effect of the pion is large at all densities.
As argued previously, this effect comes from the attractive second-order 
conribution generated by the pion potential, which is clearly quite large already at 
low density. 
As density increases, the second-order tensor contribution is reduced by the Pauli operator (and dispersion effects)
and thus retains approximately the same size. 
We also note the clear impact of the pion on the saturation density
of SNM, demonstrating the remarkable saturating effect generated by the tensor force,
particularly through the $^3S_1$ partial wave.

For neutron matter, on the other hand, the contribution
of the pion comes mostly from the (repulsive) tensor force in some major isospin-1 partial 
waves. Accordingly, 
Fig.~1(b) shows that such contribution is opposite
in sign and weaker as compared to the one in SNM, as already observed when discussing Table 
III. 
Also, the effect increases with density, in contrast to the case of SNM; see
comments in the previous paragraph.

In Table IV, we show the difference between the potential energy contributions 
to NM and SNM from the isovector mesons, as an estimate of the effect of each meson
on the potential energy part of 
 the symmetry energy. (The density is the same as in the previous tables.)
Clearly, 
in a microscopic, meson-theoretic approach the impact of the pion on the symmetry energy is the largest. 
We find this to be a point of considerable interest, since mean field theories are 
generally pionless. This is 
because the bulk of the attraction-repulsion balance needed
for a realistic description of nuclear matter can be technically                 
obtained from 
$\sigma$ and $\omega$ only, an observation that is at the very foundation
of Walecka models such as QHD-I \cite{Wal74}. 
However, in any fundamental theory of nuclear forces, the pion is the most 
important ingredient. Chiral symmetry is spontaneously broken in low-energy QCD
and the pion emerges as the Goldstone boson of this symmetry breaking \cite{ME11}.
Moreover, NN scattering data cannot be described without the pion, which is also 
absolutely crucial for the two-nucleon bound state, the deuteron. 

When moving to nuclear matter (and regardless the possibility of 
obtaining realistic values of its bulk properties, including the symmetry 
energy, with a pionless theory), this conceptual problem is not removed. 
{\it Isospin dependence is carried by the isovector mesons: Because of their 
isovector nature, these mesons contribute differently in different partial waves thus  
giving rise to isospin dependence.} (This is not the case with 
isoscalar mesons, which tend to contribute similarly in all partial waves.)
Thus, an important aspect of the physics is missing in a 
discussion of isospin dependence that does not include the pion.               
Also,                 
conclusions concerning the effect of other mesons                                  
(particularly $\rho$ and $\delta$) may be distorted due to the absence of the pion.
This may include, for instance, observations concerning 
isospin-sensitive quantities such as the neutron-proton
mass splitting in neutron-rich matter. 

As mentioned earlier, investigations of $\rho$ and $\delta$ contributions to the 
potential symmetry energy have been reported, such as the one in Refs.~\cite{DiToro,Liu}.
In Fig.~6-1 of Ref.~\cite{DiToro}, for instance, those contributions are shown 
to be very large in size (about -40 MeV and 50 MeV at saturation density for 
$\delta$ and $\rho$, respectively). Thus, the interplay between $\rho$ and $\delta$ 
is described as the equivalent, in the isovector channel, of the $\sigma$-$\omega$
interplay in the isoscalar channel \cite{Liu}. 

The dramatic differences between those and our present observations                          
originate from several sources, 
which include: 
The absence of the pion; the nature of the $\rho$ coupling; the fact that 
our meson contributions, when iterated, are reduced by the effect of the Pauli
projector. 
As mentioned previously, the role of the $\delta$ is important although subtle,
and it is found in its different contributions to I=1 and I=0 partial waves,
especially the S-waves.

In Fig.~2 we show the density dependence of the symmetry energy with 
Bonn A, B, and C. The potential model dependence comes almost 
entirely from differences among predictions of the SNM energy. 
With the three sets of predictions, we mean to estimate 
the uncertainty to be expected when using different parametrizations 
for the isovector mesons, while respecting 
the free-space NN data. 

 Figure 3 displays the momentum dependence of the 
single-proton and single-nucleon potentials in IANM, as predicted by the three
potentials. Differences are small, at most 10\% at the lowest momenta. 
We recall that the gradient between the potentials shown in Fig.~2, closely related to
the isovector optical potential, 
is the crucial mechanism that separates proton and neutron dynamics in IANM.

In closing this section, 
we take note of Ref.~\cite{Li11}, where the effect of the short-range tensor 
interaction on the symmetry energy is examined using an approximate    
expression for the second-order tensor contribution \cite{BM94}. 
It must be noted, though, that 
the variations performed 
on the short-range tensor interaction in Ref.~\cite{Li11} are 
unconstrained and, thus, to some extent arbitrary. 
\begin{figure}[!t] 
\centering 
\vspace*{-1.0cm}
\hspace*{-1.0cm}
\scalebox{0.40}{\includegraphics{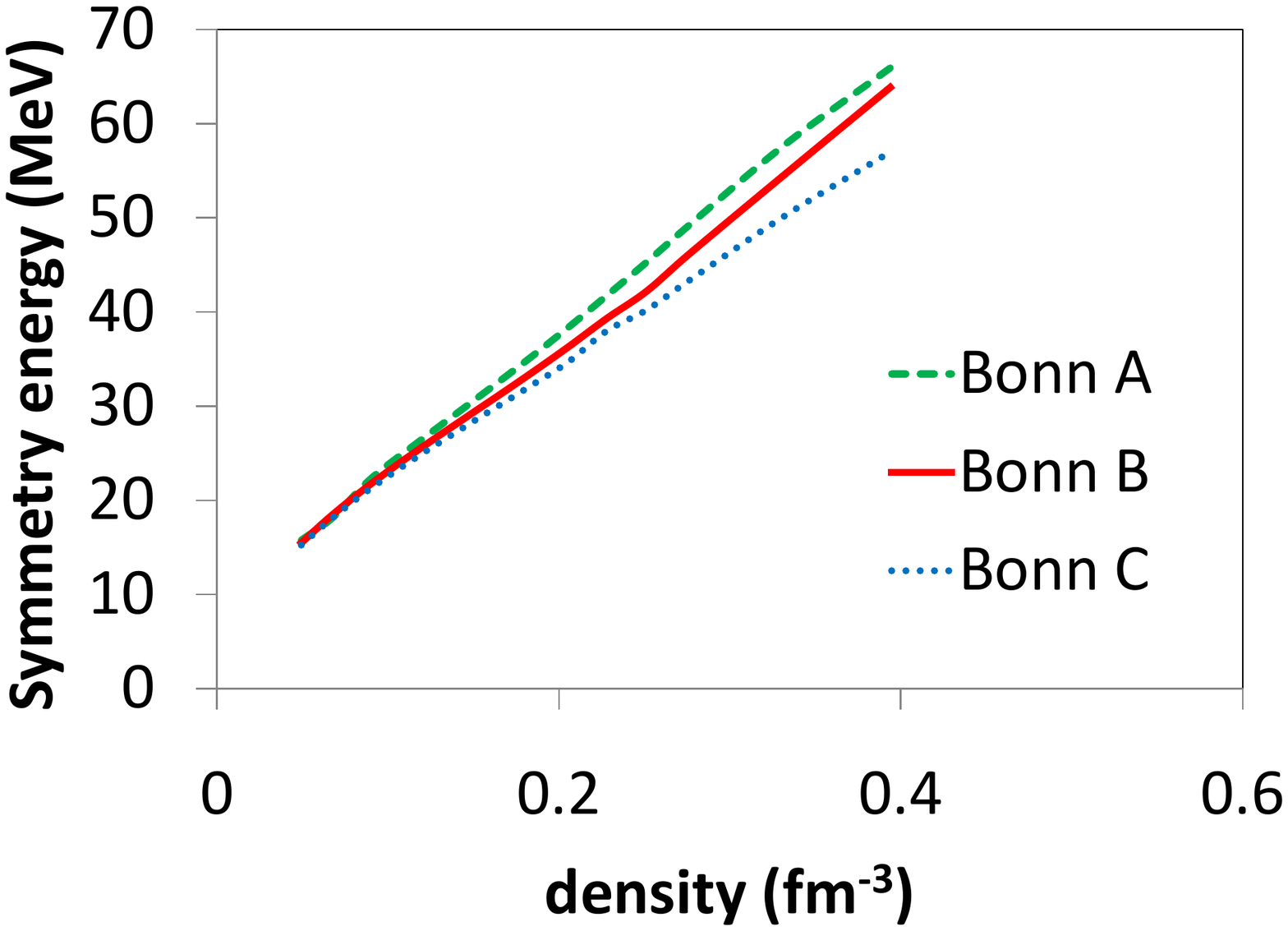}} 
\vspace*{-0.58cm}
\caption{(color online)                                        
The symmetry energy as predicted with Bonn A, B, and C.
} 
\label{two}
\end{figure}

\begin{figure}[!t] 
\centering 
\vspace*{-1.0cm}
\hspace*{-1.0cm}
\scalebox{0.50}{\includegraphics{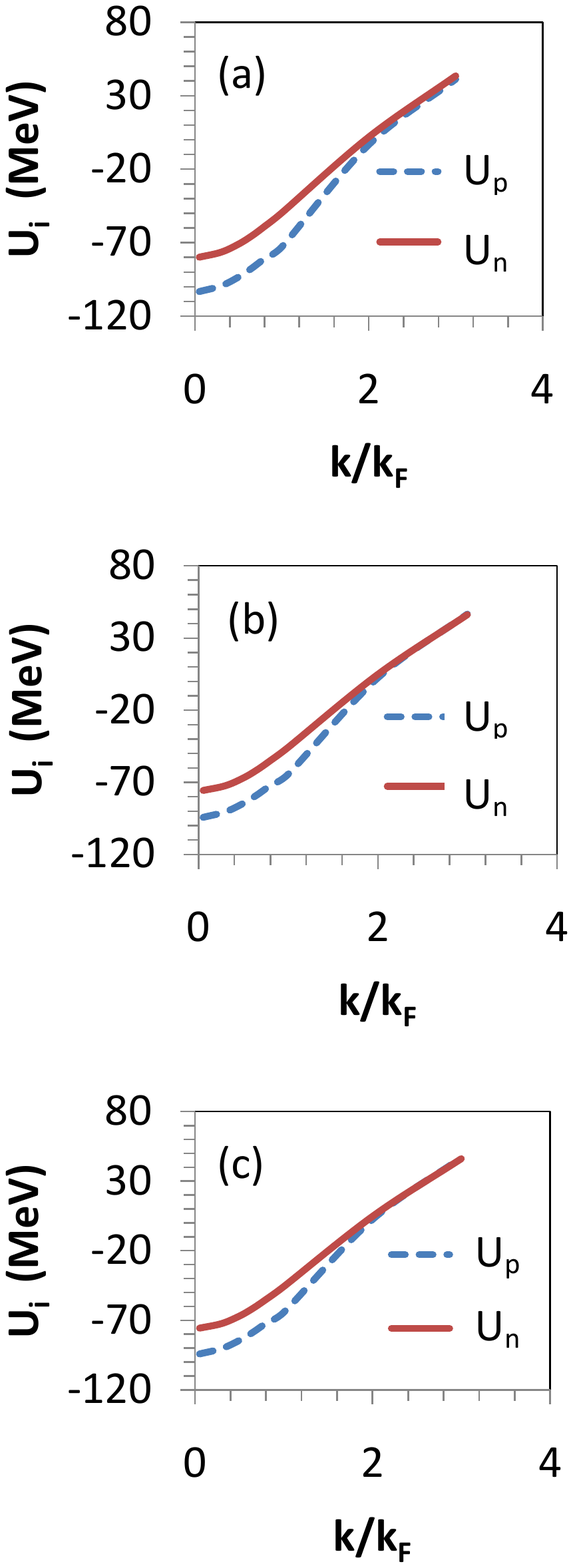}} 
\vspace*{-0.6cm}
\caption{(color online)                                        
Momentum dependence of the single-nucleon potentials in IANM, $U_i$ (i=p,n),
predicted with Bonn A, B, and C.                                    
The total density is equal to 0.185 fm$^{-3}$ and the isospin asymmetry parameter
is 0.4. The momentum is given in units of the Fermi momentum, which is equal to 
1.4 fm$^{-1}$. 
} 
\label{three}
\end{figure}

\section{Conclusions}                                                                  
We have examined the effect of the isovector mesons on the difference between the 
potential 
energies of pure neutron matter and symmetric matter. 
Our findings are easily understood in terms of the contributions of each meson
to the appropriate component of the nuclear force and the isospin dependence
naturally generated by isovector mesons. 

We find that the pion gives the largest contribution to this difference.                     
The contribution of the pion is often ovelooked, possibly because this meson
is missing from some mean field models, which are popular among users of equations 
of state. It is our opinion that conclusions regarding the interplay of 
$\rho$ and $\delta$ in phenomenological models must be taken with caution. 
                                                                      
We comment on fundamental differences between our approach and the one of 
mean field models, particularly pionless QHD theories. First, these 
differences are of conceptual relevance, since free-space NN                    
scattering and bound state are, essentially, pion physics. Furthermore, 
they can impact in a considerable way conclusions with regard to 
isospin dependent systems/phenomena. 
In order to have a fundamental basis, 
a microscopic theory of the nuclear many-body problem has to start from the bare   
NN interaction with all its ingredients.

\section*{Acknowledgments}
Support from the U.S. Department of Energy under Grant No. DE-FG02-03ER41270 is 
acknowledged.                                                                           
%\newpage

\end{document}